\documentclass[traditabstract]{aa} 
\usepackage{graphicx}
\usepackage{txfonts}
\usepackage{natbib}
\usepackage{amsmath}
\usepackage{lscape}

\usepackage[version=4]{mhchem} 

\bibpunct[]{(}{)}{;}{a}{}{,}

\begin{document}

   \title{Deuterated methyl mercaptan (CH$_3$SD): Laboratory rotational spectroscopy 
          and search toward IRAS 16293$-$2422~B\thanks{The input and output files of the fit are available as text files at CDS 
          via anonymous ftp to cdsarc.u-strasbg.fr (130.79.128.5) or via http://cdsweb.u-strasbg.fr/cgi-bin/qcat?J/A+A/621/A114}}

   \author{Olena Zakharenko\inst{1}
           \and
           Frank Lewen\inst{1}
           \and
           Vadim V. Ilyushin\inst{2,3}
           \and
           Maria N. Drozdovskaya\inst{4}
           \and
           Jes K. J{\o}rgensen\inst{5}
           \and
           Stephan Schlemmer\inst{1}
           \and
           Holger S.~P. M{\"u}ller\inst{1}
           }

   \institute{I.~Physikalisches Institut, Universit{\"a}t zu K{\"o}ln,
              Z{\"u}lpicher Str. 77, 50937 K{\"o}ln, Germany\\
              \email{zakharenko@ph1.uni-koeln.de, hspm@ph1.uni-koeln.de}
              \and
              Institute of Radio Astronomy of NASU, Mystetstv 4, 61002 Kharkiv, Ukraine 
              \and
              Quantum Radiophysics Department, V.N. Karazin Kharkiv National University, Svobody Square 4, 61022 Kharkov, Ukraine
              \and
              Center for Space and Habitability, Universit{\"a}t Bern, Sidlerstrasse 5, 3012 Bern, Switzerland
              \and
              Centre for Star and Planet Formation, Natural History Museum of Denmark, 
              University of Copenhagen, {\O}ster Voldgade 5$-$7, 1350 Copenhagen K, Denmark
              }

   \date{Received 19 October 2018 / Accepted 10 November 2018}

  \abstract
{Methyl mercaptan (also known as methanethiol), \ce{CH3SH}, has been found in the warm 
and dense parts of high- as well as low- mass star-forming regions. The aim of the present 
study is to obtain accurate spectroscopic parameters of the S-deuterated methyl mercaptan 
\ce{CH3SD} to facilitate astronomical observations by radio telescope arrays at (sub)millimeter 
wavelengths. We have measured the rotational spectrum associated with the large-amplitude internal rotation 
of the methyl group of methyl mercaptan using an isotopically enriched sample 
in the 150$-$510~GHz frequency range using the K{\"o}ln millimeter wave spectrometer. 
The analysis of the spectra has been performed up to the second excited torsional state. 
We present modeling results of these data with the RAM36 program. 
\ce{CH3SD} was searched for, but not detected, in data from the Atacama Large 
Millimeter/submillimeter Array (ALMA) Protostellar Interferometric Line Survey (PILS) 
of the deeply embedded protostar IRAS 16293$-$2422. The derived upper limit 
corresponds to a degree of deuteration of at most $\sim$18\%.}


\authorrunning{O. Zakharenko et al.}
\titlerunning{Laboratory spectroscopic study of $\ce{CH3SD}$}

\maketitle
\hyphenation{For-schungs-ge-mein-schaft}

\section{Introduction}
\label{intro}

Sulfur(S)-bearing molecules are of great astrophysical importance since they are excellent 
tracers of early protostellar evolution \citep{Charnley1997,Buckle2003,vanderTak2003,Herpin2009}. 
Moreover, their abundance is particularly sensitive to physical and chemical evolution 
in hot cores, thus sulfur has been proposed to be a chemical clock in these regions 
\citep{Charnley1997,Hatchell1998,1998Hatchell,Wakelam2011}. 
But the systematic understanding of sulfur chemistry in massive star-forming regions 
is not yet complete, namely because of the sulfur depletion problem \citep{Ruffle1999}. 
Therefore, more observations of S-bearing species are needed to test the chemical models 
for a better understanding of the star formation process. Methyl mercaptan, also known as 
methanetiol, \ce{CH3SH}, was among the molecules detected early in space by means of 
radio-astronomy, first tentatively by \citet{Turner:1977} and confirmed subsequently 
by \citet{Linke139L}. Both observations were made toward the prolific high-mass 
star-forming region Sagittarius (Sgr) B2 close to the Galactic center. The molecule was 
observed later toward the G327.3$-$0.6 hot core, the warm and dense part of a high-mass
 star-forming region \citep{Gibb:2000}, the cold core B1 \citep{2012ApJ...759L..43C}, the Orion~KL 
hot core \citep{2014ApJ...784L...7K}, the low-mass star-forming region IRAS 16293$-$2422 
\citep{2016MNRAS.458.1859M}, and the prestellar core L1544 \citep{2018MNRAS.478.5514V}. 
There is also evidence for the presence of \ce{CH3SH} toward the protostellar object 
HH212 \citep{2017ApJ...843...27L}. Unbiased molecular line surveys carried out with the 
Atacama Large Millimeter/submillimeter Array (ALMA) toward Sgr~B2(N) \citep{Holger2016} 
and IRAS 16293$-$2422 \citep{2018Drozdovskaya} detected methanethiol at levels that 
make detection of minor isotopic species plausible. 
Studying the isotopic abundance ratios can improve our understanding of the chemical and 
physical evolution of the different parts of the interstellar medium.

The enrichment of deuterium in dense molecular clouds has been of considerable interest 
for many years \citep{1989ApJ...340..906M}. The degree of deuteration has been viewed 
as an evolutionary tracer in low-mass star-forming regions 
\citep{2005ApJ...619..379C,2007prpl.conf...47C,2018arXiv180704663C}, and this may even 
apply to high-mass star-forming regions \citep{2011A&A...529L...7F}. 
\citet{2016A&A...595A.117J} carried out the Protostellar Interferometric Line Survey (PILS) 
of the low-mass protostellar binary IRAS 16293$-$2422 with ALMA covering 329$-$363~GHz. 
The survey is particularly suitable to probe the chemical content of the hot corinos, 
the warm and dense parts of the molecular cloud surrounding the protostars, because 
of the high spatial resolution of $\sim$0.5''. 
Several deuterated species were detected for the first time, among them the mono-deuterated 
isotopomers of glycolaldehyde \citep{2016A&A...595A.117J}, DNCO and the mono-deuterated 
isotopomers of formamide \citep{2016A&A...590L...6C}, HD$^{34}$S \citep{2018Drozdovskaya}, 
D$_2^{13}$CO \citep{2018A&A...610A..54P}, HDNCN \citep{2018A&A...612A.107C}, \ce{CHD2CN} 
\citep{2018arXiv180409210C}, as well as numerous additional deuterated species 
\citep{2018arXiv180808753J}. The degree of deuteration differs from about one 
to a few percent per H atom in the molecule which was explained as potentially being caused 
by different timescales on which the molecules and possibly the specific isotopologs 
were formed.  
Other recent detections of deuterated molecules, mostly in other sources, include \ce{CH3OCH2D} \citep{2013A&A...552A.117R}, 
\ce{c-C3D2} \citep{2013ApJ...769L..19S}, NH$_3$D$^+$ \citep{2013ApJ...771L..10C}, 
\ce{l-C3HD} \citep{2016A&A...586A.110S}, DOCO$^+$ \citep{2016A&A...593A..94F}, 
DCS$^+$ \citep{2016A&A...593A..94F,2016A&A...594A.117P}, and \ce{DC7N} \citep{2018MNRAS.474.5068B}.

Thus, laboratory spectroscopy of deuterated methyl mercaptan, \ce{CH3SD}, and its potential detection 
will enable astrophysicists and astrochemists to measure the D/H abundance ratios and study deuteration of 
so far little explored sulfur-bearing species. \ce{CH3SD} as well as \ce{CH3SH} are also 
of fundamental interest because of the large amplitude internal rotation of the \ce{CH3} group 
against its framework \ce{SD} or \ce{SH}, respectively.

The \ce{CH3SH} main isotopic species was subjected to numerous studies. 
Early investigations into its rotational spectrum were carried out more than 50 years ago 
\citep{Solimene:1955,Kojima:1957,Kojima:1960}. 
The investigations were extended later into the millimeter and lower submillimeter regions 
\citep{Lees:1980,Sastry:1986,Bettens:1999} and into the terahertz (1.1$-$1.5~THz) and 
far-infrared regions (50$-$550~cm$^{-1}$) \citep{Xu:2012}. 
The last study also initiated several high-resolution infrared spectroscopic investigations, 
for example, \citep{2018JMoSp.343...18L} and references therein.

The deuterated methyl mercaptan has been studied by infrared spectroscopy in the gas, liquid, 
and solid state, as well as by Raman spectroscopy, with the purpose of eliminating some 
discrepancies in the fundamental frequency assignments of \ce{CH3SH} \citep{MAY19681605}. 
The microwave spectrum of \ce{CH3SD} has been measured in the frequency range 8$-$168~GHz. 
The rotational transitions have been analyzed in the ground and first excited torsional states 
using a fourth-order effective torsion-rotation Hamiltonian \citep{TSUNEKAWA198963}. The rms 
deviation of the fit of 0.6~MHz as well as the coverage of the $J$ quantum numbers up to ten are not 
sufficient for the search of this molecule in the dense spectra of astronomical objects. 
The aim of the present investigation is to provide reliable predictions for the 
astronomical observations at millimeter and submillimeter wavelengths. We can achieve this by 
improving the rms error of the fit and by extending the quantum number and frequency ranges, 
which will result in a refined set of spectroscopic parameters. 
Subsequently, we carried out a first search for \ce{CH3SD} 
toward IRAS 16293$-$2422~B in the PILS data.

\section{Laboratory spectroscopic details}
\label{exptl}

The sample of \ce{CH3SD} (98\% atom D) has been purchased from Sigma Aldrich. 
Despite conditioning of the cell with \ce{D2O}, we had D/H exchange during 
the measurements, and strong lines of the normal methyl 
mercaptan were observed in the spectrum (see Fig. ~\ref{fgr:QBranche1}). The measurements were 
done at room temperature at a pressure of about 2~Pa. The rotational spectra have been measured 
in the frequency range from 150 up to 510~GHz using the Cologne mm/submm wave spectrometer. 
The synthesizer Agilent E8257D followed by the VDI (Virginia Diodes, Inc.) frequency 
multiplication chain has been used as a source of the signal. The RF input frequency has been 
modulated at a frequency $f=47.8$~kHz. The modulation amplitude and frequency step have been adjusted to optimize 
the S/N ratio. Schottky diode detectors have been used to detect the output frequencies. 
The output signal from the detectors has been detected by a lock-in amplifier in $2f$ mode, 
resulting in approximately second-derivative line-shapes, with a time constant 20$-$50~ms. 
A detailed description of the spectrometers may be found in \citet{Bossa:2014,Xu:2012}. 
The estimated frequency uncertainties were 30, 50, and 100~kHz depending on the S/N ratio 
and the profile of the line shape.

\section{Spectroscopic results}
\label{lab-results}

For the analysis of the spectra the rho-axis-method and the RAM36 code \citet{ILYUSHIN201026} were chosen, which had already successfully been applied for molecules with a $C_{3v}$ 
top attached to a molecular frame of $C_s$ symmetry 
\citep{ILYUSHIN201331,2014JMoSp.295...44S}. The barrier to methyl group 
internal rotation was determined to be intermediate, 440.9~cm$^{-1}$ \citep{TSUNEKAWA:1989}, 
therefore, the rotational spectrum is complicated by splitting of the torsional energy levels 
into $A$ and $E$ substates. The RAM Hamiltonian allows to perform a joint fit of rotational stacks of levels associated with several torsional states (interactions with non-torsional vibrational modes are not included in the model) and its general expression may be written as 
\begin{gather*}
H=1/2\sum_{pqnkstl}B_{pqnkstl}[J^{2p}J_z^qJ_x^nJ_y^kp_{\alpha}^s\cos(3t\alpha)\sin(3l\alpha)+\\
\sin(3l\alpha)\cos(3t\alpha)p_{\alpha}^s J_y^kJ_x^nJ_z^qJ^{2p}]  \quad(1)  
\end{gather*}
          
\noindent where the \ce{$B$_{$pqnkstl$}} are fitting parameters; $p_\alpha$ is the angular momentum 
conjugate to the internal rotation angle $\alpha$; $J_x$, $J_y$, and $J_z$ are projections 
on the $x$, $y$, and $z$ axes of the total angular momentum $J$. A more detailed description 
of the RAM36 code can be found in \citet{ILYUSHIN201331,ILYUSHIN201026}.

\begin{figure*}
 \centering
 \includegraphics[height=9cm]{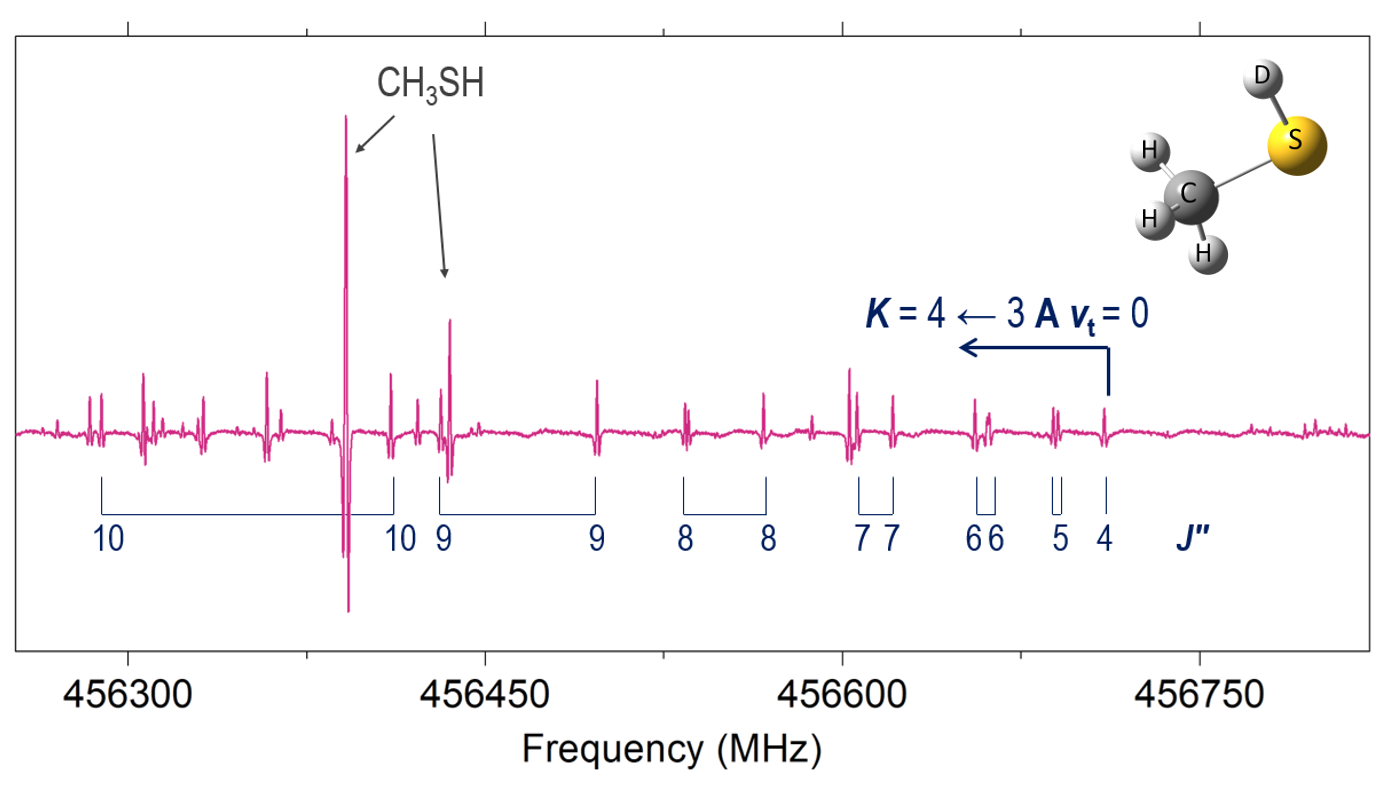}
 \caption{Part of the \ce{CH3SD} spectrum showing the \textit{$Q$} branch ($K=4\leftarrow3,~A,~\varv_t=0$). 
   Despite of preliminary condition of the cell, strong lines of the normal methyl mercaptan can be seen.}
 \label{fgr:QBranche1}
\end{figure*}

\begin{figure*}
 \centering
 \includegraphics[height=9cm]{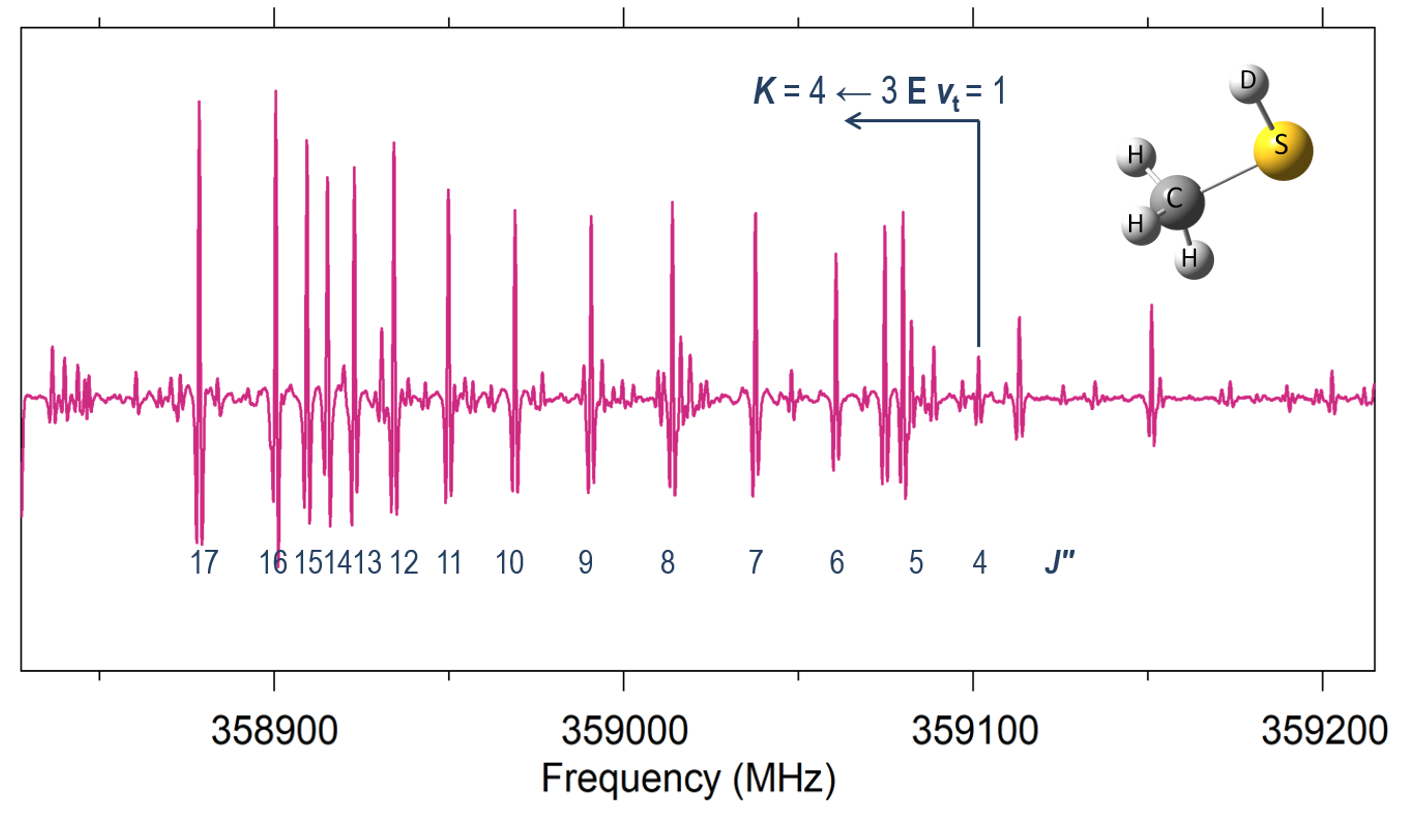}
 \caption{Part of the \ce{CH3SD} spectrum showing the \textit{$Q$} branch ($K=4\leftarrow3,~E,~\varv_t=1$). }
 \label{fgr:QBranche2}
\end{figure*}

Deuterated methyl mercaptan has two components of the electric dipole moment, $\mu_a=1.289$~D 
and $\mu_b=-0.749$~D \citep{TSUNEKAWA:1989}, so that both \textit{$a$}-type and \textit{$b$}-type 
transitions can be observed in the spectra. At the first step of our analysis, we fit the data 
from \citet{TSUNEKAWA198963} with the RAM36 code. The dipole moment components were recalculated in the RAM system. 
As it was already discussed in the literature (e.g., \citet{ILYUSHIN2003170}) in the 
molecules with large amplitude torsional motion the relative signs of the dipole moment components are important 
for obtaining correct intensity calculations. This choice should match the 
sign of $D_{ab}$ parameter, which cannot be determined from the energy level positions only. The correctness of the adopted sign choice was 
verified experimentally by comparing the relative intensities for a number of transitions which intensities are significantly affected by 
the change in relative sign of dipole moment components. Whereas in general there are two relative sign choices that match our 
experimental data, we finally adopted the choice with positive $D_{ab}$ value and relative signs of dipole moments that coincide 
with those in the main isotopologue of methyl mercaptan $\mu_a = 1.289$ D and $\mu_b = -0.749$ D. The predictions calculated 
from the initial fit allowed us to assign the $R$-branch rotational transitions with low $K_a$ quantum numbers of deuterated 
methyl mercaptan in the ground and first excited torsional states. 
Newly assigned transitions were gradually added to the dataset and a number of refinement cycles 
for the parameters were performed. The improved values of the Hamiltonian parameters provided 
reliable predictions for the higher values of the $K_a$ quantum numbers. 
The second excited torsional state has been assigned in a similar manner. 
At the next step to extend the coverage of $J$ quantum number the $Q$-, $P$-branches and 
\textit{b}-type rotational transitions have been carefully searched for and assignments have been made. 
Examples of spectral recordings showing two $Q$-branches are presented in Figs.~\ref{fgr:QBranche1} 
and \ref{fgr:QBranche2}. Asymmetry splittings in Fig.~\ref{fgr:QBranche1} become larger 
as $J$ values increase. The separation between the lines in the $Q$-branch in 
Fig.~\ref{fgr:QBranche2} decreases as $J$ values increase up to 15 and then increases. 
This effect results from the interplay of contributions of rotation and torsion motions within this torsional state.
 Finally 4905 rotational transitions 
have been assigned for the ground, first and second excited torsional states of \ce{CH3SD}, 
which, due to blending, correspond to 4434 fitted line frequencies. 
Some statistical information on the final fit is presented in Table~\ref{tbl:statisticInf}. 
The full dataset has been fit using 
78 parameters with overall weighted standard deviation 0.9. In Table~\ref{tbl:ParametersTable}, 
the final set of parameters is presented. The fits of \ce{CH3SD} and \ce{CH3SH} employ different sets of 
high order torsion-rotational parameters. Therefore, we focus our comparison on parameters of low order,
 which are given in Table~\ref{tbl:ParametersComparison}. Deuterium substitution leads to a decrease 
in the \textit{F}, $\rho$, and $V_3$ parameters as well in the rotational constant $A$. 
A similar change in these parameters is traced in deuterated methanol by \citet{WALSH200060}. 
One can notice as well a change in sign of the $D_{ab}$ parameter. As discussed earlier, the 
sign of $D_{ab}$ parameter cannot be determined from the energy level positions, and in our work we 
adopted the sign choice which matches the relative signs of dipole moment components used for 
the main isotopologue of methyl mercaptan. The experimental check of relative intensities of 
a number of transitions showed that $\mu_a = 1.289$ D and $\mu_b = -0.749$ D sign choice corresponds to 
positive $D_{ab}$ value in the case of \ce{CH3SD} and to negative $D_{ab}$ value in the case of \ce{CH3SH}.
Further comparison of low-order parameters shows that we use two fewer fourth-order parameters than in \citep{Xu:2012}. 
Our number of fourth-order parameters (22) is consistent with the total number of determinable parameters for the fourth-order
as calculated from the difference between the total number of symmetry-allowed fourth-order Hamiltonian terms and symmetry-allowed third-order contact transformation terms (\citet{NAKAGAWA:1987}).
 The input and output files of the global fit are included in the supplementary data.      

\begin{table}

  \caption{Total number of transitions and other statistical information of \ce{CH3SD} data set.}
  \label{tbl:statisticInf}
\centering
  \begin{tabular}{rrrrr}
    \hline
    $m^a$ & $N^b$ & $K_{a,\rm{max}}$$^{c}$ & $J_{\rm{max}}$$^{d}$ & $rms^{e}$ \\  
    \hline

     0 & 1004 & 18 & 53 & 54 \\
     1 & 1072 & 18 & 51 & 47 \\
  $-$3 & 848  & 17 & 42 & 53 \\
  $-$2 & 944  & 16 & 43 & 47 \\
     3 & 514  & 14 & 36 & 57 \\
     4 & 523  & 15 & 35 & 55 \\
   
    \hline
  \end{tabular}
\tablefoot{$^{a}$ Free-rotor quantum number $m$ of the lower and upper states in the rotational transition. 
 $^{b}$ Number of rotational transitions in a given category. $^{c}$ The maximum value of $K_a$ quantum number 
 in a given category. $^{d}$ The maximum value of $J$ quantum number in a given category. $^{e}$ Root mean square in kHz.}   

\end{table}

\section{Observational results}
\label{obs_res}

The Protostellar Interferometric Line Survey (PILS; project-id: 2013.1.00278.S, PI: Jes K. 
J{\o}rgensen\footnote{http://youngstars.nbi.dk/PILS/}) is an unbiased molecular line survey 
of the Class~0 protostellar binary IRAS 16293$-$2422 carried out in Band 7 of ALMA and 
covering 329.15$-$362.90~GHz at 0.244~MHz spectral resolution. Details of the survey have been 
presented by \citet{2016A&A...595A.117J}. The binary is close-by at a distance of $\sim$141~pc 
\citep{2018A&A...614A..20D}; source~A and source~B are clearly distinguished at 
the high spatial resolution of the survey of $\sim$0.5$\arcsec$ and their separation of $\sim$5.3$\arcsec$. 
The spectral sensitivity is also very high, 7$-$10~mJy beam$^{-1}$ channel$^{-1}$ or 
4$-$5 mJy beam$^{-1}$ km~s$^{-1}$ such that line confusion is reached in 
parts of the survey. This high sensitivity is very important for searching for less 
abundant molecules or for minor isotopic species of somewhat more abundant molecules. 
Many of our analyses focused on source~B because of its smaller 
lines widths of $\sim$1~km~s$^{-1}$ compared to around 3~km~s$^{-1}$ for source~A. 
Among the most exciting results is the detection of methyl chloride toward both sources 
as the first interstellar organohalogen compound \citep{2017NatAs...1..703F} in addition to 
the numerous deuterated molecules mentioned in Section~\ref{intro}.


\begin{figure}
\centering
\includegraphics[width=8cm,angle=0]{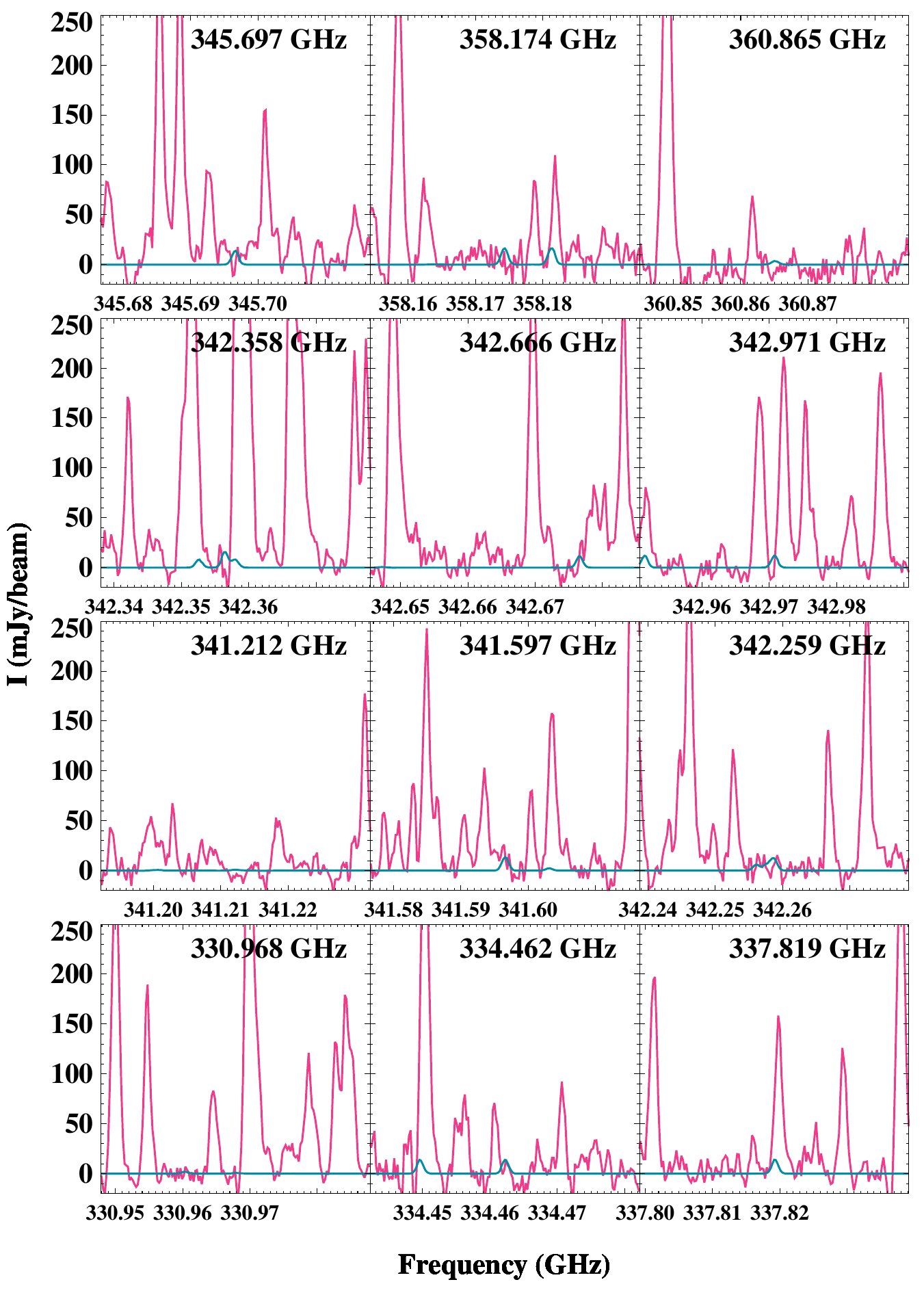}

\caption{Sections of the Protostellar Interferometric Line Survey (PILS) toward 
   a position slightly off-set from source~B displaying 
   the 12 most constraining lines of \ce{CH3SD}.}
\label{CH3SD_astro}
\end{figure}


\citet{2018Drozdovskaya} analyze column densities of several sulfur-containing molecules in 
the vicinity of source~B and compared these with data for comet 67P/Churyumov-Gerasimenko. 
A position offset by 0.5$\arcsec$ from source~B in the SW direction was used in the analysis, 
as in most of our studies of source~B. There, the effects of absorption and high 
line opacities are reduced, which could make the analyses more complicated. 
We analyzed the PILS data for lines of \ce{CH3SD} similarly to what \citet{2018Drozdovskaya} 
did for \ce{CH3SH}. We adopted an excitation temperature $T_{\rm ex} = 125$~K, 
a source size of 0.5$\arcsec$, and a full line width at half maximum (FWHM) of 1~km~s$^{-1}$. 
We did not detect \ce{CH3SD}, but we derived an upper limit on its column density of 
$\textless 8.8 \times 10^{14}$~cm$^{-2}$ ($\pm 0.88$). The 12 most constraining lines are shown 
in Fig.~\ref{CH3SD_astro}. \citet{2018Drozdovskaya} determined a column density of 
$4.8 \times 10^{15}$~cm$^{-2}$ ($\pm 0.48$) for \ce{CH3SH} \footnote{The value differs from \citet{2018Drozdovskaya} 
because the correction factor of 1.14 for the higher dust background temperature (21K) was not 
included in the paper.}. Our upper limit corresponds thus 
to a D/H ratio of less than $0.18 \pm 0.025$; the uncertainty was estimated as 
previously \citep{2018Drozdovskaya}. This value is probably not particularly 
constraining compared with the deuteration of \ce{H2CS} 
per H atom of $\sim0.05 \pm 0.007$ \citep{2018Drozdovskaya}. 
Deeper interferometric observations are required in order to conclusively verify 
whether the level of deuteration of \ce{H2CS} is lower, higher, 
or preserved at the next level of chemical complexity into \ce{CH3SH}.

\section{Conclusion and outlook}
\label{conclusion}

In the present work, the rotational spectrum of \ce{CH3SD} has been investigated in the frequency 
range 150$-$510~GHz in order to provide accurate predictions for astronomical searches. 
Extensive assignments have been made for the ground, first, and second excited torsional states 
up to high $J$ and $K_a$ quantum numbers (see Table~\ref{tbl:statisticInf}). 
The detailed modeling of the absorption spectra of the \ce{CH3SD} has been performed 
in a global fit of a dataset of 4905 rotational transitions to the RAM Hamiltonian containing 
78 parameters. The overall weighted standard deviation of the global fit is 0.9 in the range 
150 to 510 GHz, indicating that our set of parameters reproduces the assigned lines within experimental uncertainties. 
Transition frequencies calculated from these parameters should be reliable for astronomical 
observations. A first attempt to search for \ce{CH3SD} in the PILS data turned out to be 
negative. The upper limit, though not unreasonable, is not very constraining either.
Future, even more sensitive observations may provide insight into the deuteration of 
\ce{CH3SH}. A search for the potentially more abundant \ce{CH2DSH} was hampered 
by insufficient laboratory data reported for this isotopolog.

Calculations of the rotational spectrum of \ce{CH3SD} will be available in the 
catalog section of the Cologne Database for Molecular Spectroscopy, CDMS, 
\citep{2016JMoSp.327...95E}. The input and output files of the fit as well as 
auxiliary files are available in the data section of the 
CDMS\footnote{https://cdms.astro.uni-koeln.de/classic/predictions/daten/Methanethiol/}.


\begin{acknowledgements}
The work in Cologne was supported by the Deutsche Forschungsgemeinschaft (DFG) in the framework 
of the collaborative research grant SFB~956, project B3. O.Z. is funded by the DFG via the 
Ger{\"a}tezentrum ``Cologne Center for Terahertz Spectroscopy''.
The work in Kharkiv was done under support of the Volkswagen foundation. The assistance of the 
Science and Technology Center in the Ukraine is acknowledged (STCU partner project P686). 
M.N.D. acknowledges the financial support of the Center for Space and Habitability (CSH) 
Fellowship and the IAU Gruber Foundation Fellowship. 
J.K.J. acknowledges support from the European Research Council (ERC) under the European Union’s 
Horizon 2020 research and innovation program through ERC Consolidator Grant ‘‘S4F'' 
(grant agreement No 646908).
This paper makes use of the following ALMA data: ADS/JAO.ALMA\#2013.1.00278.S. 
ALMA is a partnership of ESO (representing its member states), NSF (USA) and NINS (Japan), 
together with NRC (Canada) and NSC and ASIAA (Taiwan), in cooperation with the Republic 
of Chile. The Joint ALMA Observatory is operated by ESO, AUI/NRAO and NAOJ. 
\end{acknowledgements}


\bibliographystyle{aa} 
\bibliography{bibliography} 



\newpage
\longtab{
\begin{longtable}{llll}
\caption{\label{tbl:ParametersTable} Spectroscopic parameters of \ce{CH3SD} ($cm^{-1}$)}\\
\hline\hline
$n_{tr}$\textit{$^a$} & Operator\textit{$^b$} & Par.\textit{$^{c,d}$} & \ce{CH3SD}\textit{$^e$}  \\
\hline
\endfirsthead
\caption{continued.}\\
\hline\hline
$n_{tr}$\textit{$^a$} & Operator\textit{$^b$} & Par.\textit{$^{c,d}$} & \ce{CH3SD}\textit{$^e$}  \\
\hline
\endhead
\hline
    $2_{2,0}$ & $p_\alpha^2$ & $F$ & 10.3520639(31)  \\
    $2_{2,0}$ & $(1-\cos 3\alpha)$ & $(1/2)V_3$ & 217.71250(12) \\
    $2_{1,1}$ & $p_\alpha P_a$ & $\rho$ & 0.493517098(11)  \\
    $2_{0,2}$ & $P_a^2$ & \textit{A} & 2.59513758(20)  \\
    $2_{0,2}$ & $P_b^2$ & \textit{B} & 0.42517153(13)  \\
    $2_{0,2}$ & $P_c^2$ & \textit{C} & 0.39176839(13)  \\
    $2_{0,2}$ & $(1/2)\{P_a{,}P_b\}$ &                        $2D_{ab}$ & 0.0107310(24)  \\
    $4_{4,0}$ & $(1-\cos 6\alpha)$ &                              $(1/2)V_6$ & $-0.43459(10)$  \\ 
    $4_{4,0}$ & $p_\alpha ^4$ &                          $F_m$ & $-0.38535(20)\times 10^{-3}$  \\
    $4_{3,1}$ & $p_\alpha ^3 P_a$ &                       $\rho_m$ & $-0.93969(42)\times 10^{-3}$  \\
    $4_{2,2}$ & $P^2(1-\cos 3\alpha)$ &                 $V_{3J}$ & $-0.19405784(64)\times 10^{-2}$  \\
    $4_{2,2}$ & $P_a^2(1-\cos 3\alpha)$ &               $V_{3K}$ & $0.685147(15)\times 10^{-2}$  \\ 
    $4_{2,2}$ & $(P_b^2-P_c^2)(1-\cos 3\alpha)$ & $V_{3bc}$ & $-0.15679(13)\times 10^{-3}$  \\
    $4_{2,2}$ & $(1/2)\{P_a{,}P_b\}(1-\cos 3\alpha)$ & $V_{3ab}$ & $0.887790(81)\times 10^{-2}$  \\  
    $4_{2,2}$ & $p^2_{\alpha} P^2$ &                      $F_J$ & $-0.2823067(53)\times 10^{-4}$  \\
    $4_{2,2}$ & $p^2_{\alpha} P_a^2$ &                    $F_K$ & $-0.123322(32)\times 10^{-2}$  \\  
    $4_{2,2}$ & $(1/2)p_\alpha ^2\{P_a{,}P_b\}$ &            $F_{ab}$ & $0.1561(14)\times 10^{-3}$  \\
    $4_{2,2}$ & $p^2_{\alpha}(P_b^2-P_c^2)$ &            $F_{bc}$ & $0.205539(85)\times 10^{-4}$  \\
    $4_{2,2}$ & $(1/2)\{P_a{,}P_c\}\sin 3\alpha$ &          $D_{3ac}$ & $0.22672(89)\times 10^{-1}$  \\
    $4_{2,2}$ & $(1/2)\{P_b{,}P_c\}\sin 3\alpha$ &             $D_{3bc}$ & $0.280992(53)\times 10^{-2}$  \\  
    $4_{1,3}$ & $p_\alpha P_aP^2$ &                      $\rho_J$ & $-0.369536(28)\times 10^{-4}$ \\
    $4_{1,3}$ & $p_\alpha P_a^3$ &                       $\rho_K$ & $-0.75792(11)\times 10^{-3}$ \\
    $4_{1,3}$ & $(1/2)p_\alpha \{P_a^2{,}P_b\}$ &                $\rho_{ab}$ & $0.2034(20)\times 10^{-3}$  \\  
    $4_{0,4}$ & $P^4$ &                                 $-\Delta_J$ & $0.4876778(87)\times 10^{-6}$  \\
    $4_{0,4}$ & $P^2 P_a^2$ &                            $-\Delta_{JK}$ & $0.143777(64)\times 10^{-4}$  \\
    $4_{0,4}$ & $P_a^4$ &                               $-\Delta_K$ & $0.178662(17)\times 10^{-3}$  \\
    $4_{0,4}$ & $P^2(P_b^2-P_c^2)$ &                    $-2\delta_J$ & $0.769272(19)\times 10^{-7}$  \\
    $4_{0,4}$ & $(1/2)\{P_a^2{,}(P_b^2-P_c^2)\}$ &           $-2\delta_K$ & $0.181622(70)\times 10^{-4}$  \\
    $4_{0,4}$ & $(1/2)\{P_a^3{,}P_b\}$ &                      $2D_{abK}$ & $0.5500(67)\times 10^{-4}$ \\    
    $6_{6,0}$ & $(1-\cos 9\alpha)$ &                          $(1/2)V_9$ & 0.035067(59)  \\   
    $6_{6,0}$ & $p_\alpha ^6$ &                          $F_{mm}$ & $-0.4497(47)\times 10^{-6}$  \\
    $6_{5,1}$ & $p_\alpha ^5 P_a$ &                       $\rho_{mm}$ & $-0.1252(14)\times 10^{-5}$  \\
    $6_{4,2}$ & $P^2(1-\cos 6\alpha)$ &                 $V_{6J}$ & $-0.18540(14)\times 10^{-4}$  \\         
    $6_{4,2}$ & $P_a^2(1-\cos 6\alpha)$ &               $V_{6K}$ & $-0.9114(41)\times 10^{-4}$  \\ 
    $6_{4,2}$ & $(1/2)\{P_a{,}P_b\}(1-\cos 6\alpha)$ &  $V_{6ab}$ & $-0.652(16)\times 10^{-4}$  \\ 
    $6_{4,2}$ & $(P_b^2-P_c^2)(1-\cos 6\alpha)$ & $V_{6bc}$ & $-0.2604(22)\times 10^{-4}$ \\
    $6_{4,2}$ & $(1/2)\{P_b{,}P_c\}\sin 6\alpha$ &           $D_{6bc}$ & $0.451(13)\times 10^{-4}$   \\ 
    $6_{4,2}$ & $(1/2)\{P_b{,}P_c{,}p_\alpha ^2{,}\sin 3\alpha\}$ &  $D_{3bcm}$ & $0.982(22)\times 10^{-5}$  \\ 
    $6_{4,2}$ & $p_\alpha ^4P_a^2$ &                       $F_{mK}$ & $-0.1329(18)\times 10^{-5}$  \\
    $6_{3,3}$ & $p_\alpha ^3P_aP^2$ &                      $\rho_{mJ}$ & $0.2009(39)\times 10^{-8}$  \\       
    $6_{3,3}$ & $p_\alpha ^3P_a^3$ &                       $\rho_{mK}$ & $-0.624(12)\times 10^{-6}$ \\       
    $6_{3,3}$ & $(1/2)\{P_a{,}P_b{,}P_c{,}p_\alpha{,}\sin 3\alpha\}$ & $\rho_{bc3}$ & $0.1166(18)\times 10^{-4}$  \\
    $6_{2,4}$ & $(1/2)P^2\{P_a{,}P_b\}(1-\cos 3\alpha)$ & \textit{$V_{3abJ}$} & $-0.19238(38)\times 10^{-6}$  \\  
    $6_{2,4}$ & $P^2(P_b^2-P_c^2)(1-\cos 3\alpha)$ & $V_{3bcJ}$ & $0.14478(59)\times 10^{-8}$  \\
    $6_{2,4}$ & $P^4(1-\cos 3\alpha)$ &            $V_{3JJ}$ & $ 0.4886(13)\times 10^{-8}$ \\
    $6_{2,4}$ & $P^2 P_a^2(1-\cos 3\alpha)$ &       $V_{3JK}$ & $-0.5961(43)\times 10^{-6}$ \\
    $6_{2,4}$ & $P_a^4(1-\cos 3\alpha)$ &          $V_{3KK}$ & $0.8023(43)\times 10^{-6}$ \\
    $6_{2,4}$ & $(1/2)\{P_b^2{,}P_c^2\}\cos 3\alpha$ &   $V_{3b2c2}$ & $0.3155(19)\times 10^{-7}$  \\  
    $6_{2,4}$ & $(1/2)P^2 p_\alpha ^2\{P_a{,}P_b\}$ &            $F_{abJ}$ & $0.891(72)\times 10^{-10}$ \\
    $6_{2,4}$ & $p^2_{\alpha} P^2(P_b^2-P_c^2)$ &     $F_{bcJ}$ & $-0.2107(14)\times 10^{-9}$  \\
    $6_{2,4}$ & $(1/2)p^2_{\alpha} \{P_a^2{,}(P_b^2-P_c^2)\}$ &     $F_{bcK}$ & $-0.270(11)\times 10^{-9}$  \\
    $6_{2,4}$ & $p^2_{\alpha} P^4$ &                     $F_{JJ}$ & $0.13603(79)\times 10^{-9}$  \\  
    $6_{2,4}$ & $p^2_{\alpha} P_a^2 P^2$ &                    $F_{JK}$ & $0.6308(79)\times 10^{-8}$  \\  
    $6_{2,4}$ & $p^2_{\alpha} P_a^4$ &                    $F_{KK}$ & $-0.611(46)\times 10^{-7}$  \\  
    $6_{2,4}$ & $(1/2)\{P_b^2{,}P_c^2\}p^2_{\alpha}$ &         $F_{b2c2}$ & $-0.4400(40)\times 10^{-9}$  \\
    $6_{2,4}$ & $(1/2)P^2\{P_a{,}P_c\}\sin 3\alpha$ &             $D_{3acJ}$ & $-0.367(11)\times 10^{-7}$  \\  
    $6_{2,4}$ & $(1/2)P^2\{P_b{,}P_c\}\sin 3\alpha$ &             $D_{3bcJ}$ & $-0.4477(11)\times 10^{-7}$  \\  
    $6_{2,4}$ & $(1/2)\{P_a^2{,}P_b{,}P_c\}\sin 3\alpha$ &             $D_{3bcK}$ & $0.3161(34)\times 10^{-5}$  \\  
    $6_{1,5}$ & $p_\alpha P_aP^4$ &                      $\rho_{JJ}$ & $0.9663(67)\times 10^{-10}$  \\
    $6_{1,5}$ & $p_\alpha P_a^3 P^2$ &                    $\rho_{JK}$ & $0.5270(54)\times 10^{-8}$  \\
    $6_{1,5}$ & $p_\alpha P_a^5$ &                       $\rho_{KK}$ & $0.5142(99)\times 10^{-7}$ \\
    $6_{0,6}$ & $P^6$ &                                  $\Phi_J$ & $-0.2202(13)\times 10^{-12}$  \\
    $6_{0,6}$ & $P^4 P_a^2$ &                             $\Phi_{JK}$ & $0.6867(22)\times 10^{-10}$  \\
    $6_{0,6}$ & $P^2 P_a^4$ &                             $\Phi_{KJ}$ & $0.1254(12)\times 10^{-8}$  \\
    $6_{0,6}$ & $P_a^6$ &                                $\Phi_K$ & $0.1440(11)\times 10^{-7}$  \\
    $6_{0,6}$ & $P^4(P_b^2-P_c^2)$ &                    $2\phi_J$ & $0.1469(64)\times 10^{-13}$  \\ 
    $6_{0,6}$ & $(1/2)P^2\{P_a^2{,}(P_b^2-P_c^2)\}$ &           $2\phi_{JK}$ & $0.16756(88)\times 10^{-9}$  \\    
    $8_{6,2}$ & $P^2(1-\cos 9\alpha)$ &                 $V_{9J}$ & $-0.2433(13)\times 10^{-5}$  \\  
    $8_{6,2}$ & $P_a^2(1-\cos 9\alpha)$ &                 $V_{9K}$ & $-0.1136(48)\times 10^{-4}$  \\  
    $8_{6,2}$ & $(1/2)\{P_b{,}P_c{,}p_{\alpha} ^4{,}\sin 3\alpha\}$ &  $D_{3bcmm}$ & $-0.828(29)\times 10^{-8}$  \\ 
    $8_{4,4}$ & $P^4(1-\cos 6\alpha)$ &                 $V_{6JJ}$ & $0.373(20)\times 10^{-9}$  \\         
    $8_{4,4}$ & $P^2 P_a^2(1-\cos 6\alpha)$ &                 $V_{6JK}$ & $-0.1430(11)\times 10^{-7}$ \\     
    $8_{4,4}$ & $(1/2)\{P_b^2{,}P_c^2\}\cos 6\alpha$ &   $V_{6b2c2}$ & $0.343(13)\times 10^{-8}$   \\  
    $8_{4,4}$ & $(1/2)\{P_a{,}P_b{,}P_c^2\}\cos 6\alpha$ & $V_{6abc2}$ & $-0.1371(34)\times 10^{-6}$   \\           
    $8_{4,4}$ & $P^2(P_b^2-P_c^2)(1-\cos 6\alpha)$ & $V_{6bcJ}$ & $0.9451(71)\times 10^{-9}$ \\
    $10_{6,4}$ & $P^4(1-\cos 9\alpha)$ &                 $V_{9JJ}$ & $0.954(13)\times 10^{-9}$   \\    
    $10_{6,4}$ & $(1/2)\{P_b^2{,}P_c^2\}\cos 9\alpha$ &   $V_{9b2c2}$ & $0.2778(94)\times 10^{-8}$  \\  
    $10_{3,7}$ & $(1/2)\{P_a{,}P_b^3{,}P_c^3{,}p_{\alpha}{,}\sin 3\alpha\}$ & $\rho_{3b3c3}$ & $-0.596(28)\times 10^{-13}$   \\
\hline
\end{longtable}
\tablefoot{$^{a}$ \textit{n=t+r}, where n is the total order of the operator, \textit{t} is the order of the torsional part and \textit{r} is the order of the rotational part, respectively. The ordering scheme of \citet{NAKAGAWA:1987} is used. $^{b}$ \ce{\{A,B,C,D\}} = ABCD + DCBA. \ce{\{A,B,C\}} = ABC + CBA. \ce{\{A,B\}} = AB + BA. The product of the operator in the first column of a given row and the parameter in the third column of that row gives the term actually used in the torsion-rotation Hamiltonian of the program, except for \textit{F}, $\rho$ and \textit{$A_{RAM}$}, which occur in the Hamiltonian in the form $F(p_a + \rho P_a)^2 + A_{RAM}P_a^2$. $^{c}$ Parameter nomenclature is based on the subscript procedure of \citet{XU:2008305}. $^{d}$ Values of the parameters in inverse centimeters, except for $\rho$, which is unitless. $^{e}$ Statistical uncertainties are given in parentheses as one standard uncertainty in units of the last digits. }
}


\newpage
\longtab{
\begin{longtable}{lllll}
\caption{\label{tbl:ParametersComparison} Comparison of the low-order parameters of \ce{CH3SD} and \ce{CH3SH}}\\
\hline\hline
$n_{tr}$\textit{$^a$} & Operator\textit{$^b$} & Par.\textit{$^{c,d}$} & \ce{CH3SD}\textit{$^e$} & \ce{CH3SH}\textit{$^{e,f}$} \\
\hline
\endfirsthead
\caption{continued.}\\
\hline\hline
$n_{tr}$\textit{$^a$} & Operator\textit{$^b$} & Par.\textit{$^{c,d}$} & \ce{CH3SD}\textit{$^e$} & \ce{CH3SH}\textit{$^{e,f}$} \\
\hline
\endhead
\hline
    $2_{2,0}$ & $p_\alpha^2$ & $F$ & 10.3520639(31) & 15.04020465(66) \\
    $2_{2,0}$ & $(1/2)(1-\cos 3\alpha)$ & $V_3$ & 435.42500(24) & 441.442236(10) \\
    $2_{1,1}$ & $p_\alpha P_a$ & $\rho$ & 0.493517098(11) & 0.651856026(13) \\
    $2_{0,2}$ & $P_a^2$ & \textit{A} & 2.59513758(20) & 3.42808445(84) \\
    $2_{0,2}$ & $P_b^2$ & \textit{B} & 0.42517153(13) & 0.43201954(87) \\
    $2_{0,2}$ & $P_c^2$ & \textit{C} & 0.39176839(13) & 0.41325076(83) \\
    $2_{0,2}$ & $\{P_a{,}P_b\}$ & $D_{ab}$ & 0.0053655(12) & $-0.0073126(59)$ \\
    $4_{4,0}$ & $(1/2)(1-\cos 6\alpha)$ & $V_6$ & $-0.86918(20)$ & $-0.572786(15)$ \\ 
    $4_{4,0}$ & $p_\alpha ^4$ &                          $F_m$ & $-0.38535(20)\times 10^{-3}$ & $-0.114016(10)\times 10^{-2}$ \\
    $4_{3,1}$ & $p_\alpha ^3 P_a$ &                       $\rho_m$ & $-0.93969(42)\times 10^{-3}$ & $-0.360009(28)\times 10^{-2}$ \\
    $4_{2,2}$ & $P^2(1-\cos 3\alpha)$ &                 $V_{3J}$ & $-0.19405784(64)\times 10^{-2}$ & $-0.217540(84)\times 10^{-2}$ \\
    $4_{2,2}$ & $P_a^2 (1-\cos 3\alpha)$ &               $V_{3K}$ & $0.685147(15)\times 10^{-2}$ & $0.724978(19)\times 10^{-2}$ \\ 
    $4_{2,2}$ & $(P_b^2-P_c^2)(1-\cos 3\alpha)$ & $V_{3bc}$ & $-0.15679(13)\times 10^{-3}$ & $-0.92104(47)\times 10^{-4}$ \\
    $4_{2,2}$ & $\{P_a{,}P_b\}(1-\cos 3\alpha)$ & $V_{3ab}$ & $0.443895(41)\times 10^{-2}$ & $0.61562(30)\times 10^{-2}$ \\  
    $4_{2,2}$ & $p^2_{\alpha} P^2$ &                      $F_J$ & $-0.2823067(53)\times 10^{-4}$ & $-0.8106(38)\times 10^{-4}$ \\
    $4_{2,2}$ & $p^2_{\alpha} P_a^2$ &                    $F_K$ & $-0.123322(32)\times 10^{-2}$ & $-0.483287(30)\times 10^{-2}$ \\  
    $4_{2,2}$ & $p_\alpha ^2\{P_a{,}P_b\}$ &            $F_{ab}$ & $0.7807(70)\times 10^{-4}$ & $0.843(45)\times 10^{-4}$ \\
    $4_{2,2}$ & $2p^2_{\alpha}(P_b^2-P_c^2)$ &            $F_{bc}$ & $0.102769(43)\times 10^{-4}$ & $0.0536(41)\times 10^{-4}$ \\
    $4_{2,2}$ & $\{P_a{,}P_c\}\sin 3\alpha$ &          $D_{3ac}$ & $0.11336(45)\times 10^{-1}$ & $0.1036(15)\times 10^{-1}$ \\
    $4_{2,2}$ & $\{P_b{,}P_c\}\sin 3\alpha$ &             $D_{3bc}$ & $0.140496(27)\times 10^{-2}$ & $0.665(14)\times 10^{-3}$ \\  
    $4_{1,3}$ & $p_\alpha P_a P^2$ &                      $\rho_J$ & $-0.369536(28)\times 10^{-4}$ & $-0.4726(54)\times 10^{-4}$ \\
    $4_{1,3}$ & $p_\alpha P_a^3$ &                       $\rho_K$ & $-0.75792(11)\times 10^{-3}$ & $-0.30381(74)\times 10^{-2}$ \\
    $4_{1,3}$ & $p_\alpha \{P_a^2{,}P_b\}$ &                $\rho_{ab}$ & $0.1017(10)\times 10^{-3}$ & $0.999(67)\times 10^{-4}$ \\  
   $4_{1,3}$ & $p_\alpha \{P_a{,}(P_b^2-P_c^2)\}$ &            $\rho_{bc}$ & $ - $ & $-0.0462(39)\times 10^{-4}$ \\
    $4_{0,4}$ & $-P^4$ &                                 $\Delta_J$ & $0.4876778(87)\times 10^{-6}$ & $0.538140(23)\times 10^{-6}$ \\
    $4_{0,4}$ & $-P^2 P_a^2$ &                            $\Delta_{JK}$ & $0.143777(64)\times 10^{-4}$ & $-0.066(26)\times 10^{-5}$ \\
    $4_{0,4}$ & $-P_a^4$ &                               $\Delta_K$ & $0.178662(17)\times 10^{-3}$ & $0.7425(48)\times 10^{-3}$ \\
    $4_{0,4}$ & $-2P^2(P_b^2-P_c^2)$ &                    $\delta_J$ & $0.384636(10)\times 10^{-7}$ & $0.224788(88)\times 10^{-7}$ \\
    $4_{0,4}$ & $-\{P_a^2{,}(P_b^2-P_c^2)\}$ &           $\delta_K$ & $0.090811(35)\times 10^{-4}$ & $0.10483(32)\times 10^{-4}$ \\
    $4_{0,4}$ & $P^2\{P_a{,}P_b\}$ &                      $D_{abJ}$ & $ - $ & $-0.956(60)\times 10^{-7}$ \\    
    $4_{0,4}$ & $\{P_a^3{,}P_b\}$ &                      $D_{abK}$ & $0.2750(34)\times 10^{-4}$ & $0.202(23)\times 10^{-4}$ \\    
     &  &                 $\theta_{RAM}$ & 0.14$^\circ$ & $-$0.14$^\circ$ \\      
\hline
\end{longtable}
\tablefoot{$^{a}$ \textit{n=t+r}, where n is the total order of the operator, \textit{t} is the order of the torsional part and \textit{r} is the order of the rotational part, respectively. The ordering scheme of \citet{NAKAGAWA:1987} is used. $^{b}$ \ce{\{A,B,C,D\}} = ABCD + DCBA. \ce{\{A,B,C\}} = ABC + CBA. \ce{\{A,B\}} = AB + BA. The product of the operator in the first column of a given row and the parameter in the third column of that row gives the term actually used in the torsion-rotation Hamiltonian of the program, except for \textit{F}, $\rho$ and \textit{$A_{RAM}$}, which occur in the Hamiltonian in the form $F(p_a + \rho P_a)^2 + A_{RAM}P_a^2$. $^{c}$ Parameter nomenclature is based on the subscript procedure of \citet{XU:2008305}. $^{d}$ Values of the parameters in inverse centimeters, except for $\rho$, which is unitless, and for \textit{$\theta_{RAM}$} in degrees. $^{e}$ Statistical uncertainties are given in parentheses as one standard uncertainty in units of the last digits. $^{f}$ Not all the parameters used for the analysis in \citet{Xu:2012} listed here. }
}

\end{document}